\documentclass[twocolumn,showpacs,preprintnumbers,amsmath,amssymb]{revtex4}
\usepackage{graphicx}% Include figure files
\usepackage{dcolumn}% Align table columns on decimal point
\usepackage{bm}% bold math

\begin{document}

\preprint{APS/123-QED}

\title{Hidden symmetry and nonlinear paraxial atom optics.}

\author{Fran\c{c}ois Impens}

\affiliation{SYRTE, Observatoire de Paris, CNRS, 61 Avenue
de l'Observatoire, 75014 Paris, France}

\affiliation{and Instituto de Fisica, Universidade Federal do
Rio de Janeiro. Caixa Postal 68528, 21941-972 Rio de Janeiro, RJ,
Brazil}

\date{\today}

\begin{abstract}
A hidden symmetry of the nonlinear wave equation is exploited to
analyze the propagation of paraxial and uniform atom-laser beams in
time-independent and quadratic transverse potentials with cylindrical symmetry. The quality factor and the paraxial $ABCD$
formalism are generalized to account exactly for mean-field
interaction effects in such beams. Using an approach based on moments, these theoretical tools provide a very simple and yet exact picture of the interacting beam profile evolution. Guided atom laser experiments are discussed. This treatment addresses simultaneously optical and atomic beams in a unified manner, exploiting the formal analogy between nonlinear optics and nonlinear paraxial atom optics.
\end{abstract}

\maketitle

\section{Introduction.}

The realization of quasi-continuous atom lasers, building upon a
great theoretical~\cite{AtomLaserTheory} and
experimental~\cite{AtomLaserExp,LeCoq01,Guerin06,David08a} effort,
is a topic of considerable excitement: beyond the challenge of
producing a fully coherent atomic beam, such devices appear indeed
as promising tools for various applications, ranging from
precision experiments using atom interferometry~\cite{Interfero}
to the investigation of basic physical phenomena such as resonant
nonlinear~\cite{Tobias} quantum transport~\cite{LeboeufPavloff}.
While the first atom lasers involved a matter-wave beam outcoupled
into free space~\cite{AtomLaserExp} and experiencing a free fall,
recent experiments have demonstrated the possibility to obtain a
guided atomic beam subject to an acceleration several orders of
magnitude smaller than gravity~\cite{Guerin06,David08a}. In such
beams the atomic flux and longitudinal velocity are
well-monitored, and the beam linear density - i.e. the number of
atoms per unit length - is constant on the propagation axis with a
good approximation. Interaction effects on the propagation of the
extracted atomic beam are usually considered as negligible outside
the source condensate, and as such are discarded from the analysis.
Using a description of the atomic beam based on
moments~\cite{DavidMoments}, it is in fact possible to account
exactly for mean-field interaction effects in such guided atom
lasers, provided that the atoms propagate at a constant velocity
and in suitable transverse potentials. The purpose of this article
is to expose this method, exploiting commutation relations
reminiscent of a hidden symmetry of the 2D Schr\"odinger
equation~\cite{KaganPitaev,Pitaevskii97}. I explore the
connection of previous results of nonlinear
optics~\cite{Vlasov71,Pare92,Porras93} to this symmetry
 and I extend them to nonlinear
atom optics, giving a compact proof valid simultaneously for both
optical and atomic beams.

The similarity between the equations of propagation for light and
atomic waves~\cite{BordeHouches,Lenz93}, enabling one to reproduce
with matter waves many nonlinear optical
phenomena~\cite{NonlinearAtomexp}, has already been used to adapt
successful theoretical methods from optics to atom
optics~\cite{MeystreBook}. Concerning the linear propagation of
matter waves, a relevant example is the $ABCD$ matrix
formalism~\cite{Kogelnik65} introduced by Bord\'e for atomic
clouds~\cite{BordeABCD}. This approach computes very efficiently
the propagation of dilute atomic wave-packets in time-dependent
quadratic potentials. Among other examples, let us mention the
introduction of a paraxial wave equation and of a paraxial $ABCD$
formalism for atomic beams~\cite{LeCoq01,Riou08}, the
characterization of a multiple-mode atomic beam with a quality
factor~\cite{Riou06,Riou08,Impens08a}, and the treatments of
atomic interactions with a lensing term~\cite{LeCoq01,Busch02}.
Although the optical approach of guided atom laser propagation is
not exclusive - other treatments rely on hydrodynamic
equations~\cite{DavidGuidedPropagation} -, it provides a reliable
and tractable characterization of such beams.

 In this spirit, I considered two general results of nonlinear optics concerning the transverse profile of paraxial
light beams. In a uniform medium, the second-order moment of the
transverse intensity distribution follows a parabolic law, even in the
presence of a Kerr effect~\cite{Vlasov71}. In a graded-index
medium, the transverse width oscillates with a period independent
from the strength of the nonlinearity~\cite{Pare92}. These
results, which attest a universal behavior of the beam width
evolution, are indeed intimately connected to a hidden symmetry~\cite{KaganPitaev,Pitaevskii97,Ghosh01} of
the paraxial equation. They give a simple and yet exact expression of the transverse size
of light beams propagating in a nonlinear medium and have had
several applications in nonlinear
optics~\cite{Porras93,AppliNonlinOptics}. As suggested in
Refs.~\cite{Riou06,Riou08}, these properties could be also
relevant for nonlinear atom optics.

 Here, I use them to develop a simple $ABCD$ matrix formalism,
 suitable to address the propagation of interacting atom-laser beams satisfying the following
 assumptions: the beam is stationary, paraxial, of constant linear density along the propagation axis, and it
 propagates in a transverse potential which is time-independent, quadratic
 and of cylindrical symmetry. As a preliminary step, we will first analyse the free
 expansion of such interacting atom laser beams and discuss
 experimental results~\cite{David08a}.

\section{Equations of propagation for paraxial light and atomic waves in a non-linear medium.}
\label{sec:equations of propagation}

In this section, we show that a
 paraxial light wave in a transparent and isotropic nonlinear
 medium and an interacting paraxial atomic beam
of constant velocity follow formally equivalent equations of propagation, which take the form of a 2D Schr\"odinger
equation
\begin{equation}
 2 i k \frac {\partial \psi_{\bot}}
{\partial u} = - \epsilon \Delta_T \psi_{\bot}  + \gamma
|\psi_{\bot}|^2 \psi_{\bot} +\epsilon k^2 \alpha^2(u) r^2
\psi_{\bot} \,. \nonumber \\
\end{equation}
 This similarity between the propagation of light and
atomic waves had been pointed out, in the linear regime, by
Bord\'e~\cite{BordeHouches}. A formal analogy between
mean-field atomic interactions and the Kerr effect is also
discussed in Ref.~\cite{Lenz93}.

\subsection{Equation of propagation for optical waves.}
\label{subsec:equation paraxial optics}

 We consider a paraxial and monochromatic light wave propagating in a non-absorptive
and nonlinear medium of graded index (GRIN). Specifically, its
linear relative permittivity is assumed to have a quadratic
position-dependence of the form
$\epsilon_r(x,y,z)=\epsilon^0_{r}(1- \beta^2(z) r^2)$. This yields
a medium index $n(r,z)=n_0(1-\beta^2(z) r^2)^{1/2}$, quadratic in
the vicinity of the propagation axis where $r \ll \beta^{-1}(z)$.
The index is greater on the propagation axis so that the medium
acts as a converging lens and traps rays around this axis. The
electric field is assumed monochromatic, written in the form
\begin{equation}
\label{eq:paraxial electric field definition}
 \mathbf{E}(x,y,z,t)=   \psi_{\bot}(x,y,z) e^{i (k z
- \omega t)} \mathbf{a} \,, \nonumber
\end{equation}
and it satisfies the paraxial approximation, i.e. $\left| \partial
\psi_{\bot} /
\partial z \right| \ll k |\psi_{\bot}|$. Using this approximation
in Maxwell's equation, one obtains the nonlinear paraxial equation
for optical waves~\cite{BookNonLinOptics}
\begin{equation}
\label{eq:paraxial equation optics} 2 i k \frac {\partial
\psi_{\bot}} {\partial z} = \Delta_T \psi_{\bot}  + \gamma
|\psi_{\bot}|^2 \psi_{\bot} - k^2 \beta^2 r^2 \psi_{\bot} \:.
\end{equation}
The wave-vector $k$ satisfies the dispersion relation
$k^2=\epsilon^0_r \omega^2 /c^2$, and the nonlinear coefficient
$\gamma=\chi_e^{(3)} \omega^2 /c^2$ is proportional to the
nonlinear susceptibility $\chi_e^{(3)}$ at the frequency $\omega$.
$\Delta_T$ is the transverse Laplacian acting on the coordinates
$x,y$. This equation appears as a 2D nonlinear Schr\"odinger
equation in which the coordinate $z$ along the propagation axis
plays the role usually devoted to the time in quantum
mechanics~\cite{BordeHouches}.

\subsection{Equation of propagation for atomic waves.}
\label{subsec:equation paraxial atom optics}

We give a reminder on the propagation equation for a paraxial and
monochromatic atom laser evolving in the stationary regime within
an external potential~\cite{LeCoq01,Riou08}. The propagation
equation is derived in the general case, but the results presented
later apply only in the absence of a longitudinal potential.

Let us consider a wave-function $\psi$, solution of the
time-independent Schr\"odinger equation
\begin{equation}
\hat{H} \psi(\mathbf{r})=E \psi(\mathbf{r}) \nonumber \,.
\end{equation}
 The Hamiltonian $\hat{H}$
accounts for interactions treated in the mean-field approximation
and for an external potential $U$, sum of a transverse cylindrical
potential $U_{\perp}(r,z)= m/2 \times \omega_{\perp}^2(z) \: r^2$
and of a
 longitudinal one $U_{\!/\!/}(z)$. As in Ref.~\cite{Jackson03}, the wave-function is factorized in a 2D+1D decomposition as
\begin{equation}
\psi(x,y,z)= \psi_{\perp}(x,y,z) \, \psi_{\!/\!/}(z). \nonumber
\end{equation}
In our conventions, the transverse wave-function is normalized to
unity $\forall z \: \: \int {\rm d} x {\rm d} y
|\psi_{\perp}|^2(x,y,z)=1$. The longitudinal wave-function
$\psi_{\!/\!/}$ verifies the 1D time-independent Schr\"odinger
equation
\begin{equation}
-\frac{\hbar^2}{2m}\frac{\partial^2 \psi_{\!/\!/}}{\partial z^2}
+U_{/\!/}\psi_{\!/\!/}=E \psi_{\!/\!/}\,, \label{eq:Schro1D}
\nonumber
\end{equation}
 which can be solved with the WKB method as
\begin{equation}
\psi_{\!/\!/}(z)= \sqrt{\frac {m \mathcal{F}} {p(z)}}
\exp\!\left[\frac{i}{\hbar}\int_{z_0}^z \mathrm{d}u\,p(u) \right]
\label{eq:solWKB1D} \nonumber \,.
\end{equation}
 $\mathcal{F}$ is the
matter-wave flux through any transverse plane,
$p(z)=\sqrt{2m\left(E - U_{/\!/}(z)\right)}$ is the classical
momentum along $z$, and the integral in the exponential argument
starts at an arbitrary coordinate $z_0$ of the propagation axis
corresponding to a point within the considered propagation zone.

For a beam propagating in the classically allowed region where
$U_{/\!/}(z)<E$, such WKB approach is indeed valid only if the
wave-vector $k(z)=p(z)/\hbar$ satisfies the condition $|{\rm d} k/
{\rm d} z|/k \ll k$, otherwise quantum reflections may occur.
Equivalently, the longitudinal potential $U_{/\!/}$ should vary
smoothly enough to verify $|{\rm d} U_{/\!/}/ {\rm d} z| \ll
\sqrt{8 m} (E-U_{/\!/})^{3/2} / \hbar $.

In a paraxial beam, the average momenta $\langle p_{\bot x,y,z}
\rangle$ of the transverse wave-function $\psi_{\perp}$ must
satisfy $|\langle p_{\bot x,y,z} \rangle| \ll p(z)$. The second
derivative of the wave-function $\psi_{\perp}$ on the coordinate
$z$ is then negligible, and the wave-function $\psi_{\perp}$
satisfies
\begin{equation}
\label{eq:atomlaserparaxial0} \left[i\hbar \frac {p(z)} {m} \frac
{\partial} {\partial z} + \frac {\hbar^2}{2m} \Delta_T - \frac {4
\pi \hbar^2 a_s} {m} |\psi_{/\!/}|^2 |\psi_{\perp}|^2 - U_\perp
\right] \psi_{\perp}=0
\end{equation}
with $a_s$ the s-wave atomic scattering length.

At this point, we note a significant difference between the
nonlinear optical and atom-optical propagations. In typical
experimental conditions in optics, the refraction index and the
longitudinal wave-vector of the beam are constant along the
propagation axis with a good approximation. In contrast,
atom-laser experiments can involve a longitudinal potential which
may change significantly the longitudinal atomic momentum during
the propagation. This induces a difference between the optical
propagation equation~\eqref{eq:paraxial equation optics} and the
atomic propagation equation~\eqref{eq:atomlaserparaxial0}: in the
latter, the coefficients of the first-order derivative and of the
nonlinear term become $z$-dependent in presence of a longitudinal
potential. One can eliminate this dependence in the first-order
derivative term by performing the following variable change,
replacing the longitudinal coordinate $z$ by the parameter
\begin{equation}
\label{eq:definition zeta}
 \tau(z) = \int_{z_0}^{z}\! {\rm d} z\, \frac {m} {p(z)} \,,
 \nonumber
\end{equation}
defined as the time needed classically to propagate from the axis point of coordinate $z_0$ to the point of coordinate $z$.
Eq.\eqref{eq:atomlaserparaxial0} takes then the form
\begin{equation}
\label{eq:atomlaserparaxial1} 2 i k \frac {\partial \psi_{\perp}}
{\partial \tau} =  - \Delta_T\psi_{\perp}  +  8 \pi
 a_s n_{1D}(\tau)  |\psi_{\perp}|^2 \psi_{\perp} + k^2 \omega^2_{\perp}  r^2
\psi_{\perp} \,.
\end{equation}
We noted $k=m/\hbar$ and introduced the linear atomic density
$n_{1D}=|\psi_{/\!/}|^2$. For a dilute atomic beam, this equation
is formally equivalent to the optical equation~\eqref{eq:paraxial
equation optics}.

 For an interacting atomic beam, however, the
coefficient of the nonlinear term in
Eq.\eqref{eq:atomlaserparaxial1} still depends on the propagation
parameter $\tau$. In the WKB treatment where the linear density
reads $n_{1D}(\tau)=m \mathcal{F}/p(\tau)$, this coefficient is a
constant only if no significant longitudinal potential is applied
to the atomic beam, which propagates then with a constant
longitudinal velocity. The nonlinear optical
equation~\eqref{eq:paraxial equation optics} and atom-optical
equation~\eqref{eq:atomlaserparaxial1} are then fully equivalent.

\subsection{Quantum mechanical interpretation of the nonlinear wave equation.}

Both equations of propagation can be written in the generic form
\begin{equation}
\label{eq:paraxial equation} 2 i k \frac {\partial \psi_{\bot}}
{\partial u} = - \epsilon \Delta_T \psi_{\bot}  + \gamma
|\psi_{\bot}|^2 \psi_{\bot} +\epsilon k^2 \alpha^2(u) r^2
\psi_{\bot} \,.
\end{equation}
with $u=z$, $k= \sqrt{\epsilon^0_r} \omega /c$, $\epsilon=-1$,
$\gamma=\chi_e^{(3)} k^2/ \epsilon^0_r$ and $\alpha=\beta$ for the
optical equation, and $u=\tau$, $k= m/\hbar$, $\epsilon=1$,
$\gamma=8 \pi
  n_{1D} a_s$ and $\alpha=\omega_{\perp}$ for the atomic
 equation.

One must assume that no longitudinal potential is driving the atoms in order to obtain a nonlinear coefficient $\gamma$ independent
 of the coordinate $\tau$ in the atomic equation. As stated previously, we make this assumption for the subsequent developments: otherwise, the equations of motion for the beam moments are not solved by the $ABCD$ matrix approach presented here. However, the coefficient $\alpha$ of the quadratic effective potential may depend on the coordinate $u$ in both the optical and atomic equations.

If one makes the correspondence $2k \rightarrow \hbar$ and $u
\rightarrow t$, Eq.\eqref{eq:paraxial equation} appears as the
Schr\"odinger equation~\footnotemark[1] \footnotetext[1]{To treat
atomic beams only, one could indeed obtain directly a 2D
Schr\"odinger equation from Eq.\eqref{eq:atomlaserparaxial0}
instead of Eq.\eqref{eq:paraxial equation}, without having to do
the correspondence $2 k \rightarrow \hbar$. The great advantage of
starting from Eq.\eqref{eq:paraxial equation} is that one obtains
a treatment addressing simultaneously optical and atomic beams.
This also shows the insight provided by a quantum-mechanical
approach of the optical paraxial equation.} of a 2D
wave-function~\cite{BordeHouches}. The corresponding Hamiltonian
\begin{equation}
\label{eq:paraxial Hamiltonian}
 \hat{H}=\epsilon \hat{K}+\hat{V}+\epsilon \hat{U} \,,
\end{equation}
involves three terms
\begin{equation}
\label{eq:paraxial Hamiltonian terms} \left\{
\begin{aligned}  & \hat{K}= \hat{p}^2 / \hbar^2 \\
& \hat{V}= \frac 1 2 \gamma \: \delta^{(2)}
(\mathbf{r}-\mathbf{r}')   \\
& \hat{U}=k^2 \alpha^2 \hat{r}^2 \,.
\end{aligned}
\right.
\end{equation}
 We have introduced the operator $\hat{\mathbf{p}}= - i \hbar (\nabla_x, \nabla_y)$ satisfying
$[\hat{x},\hat{p}_x]=i \hbar$, associated with the transverse
momentum of the atoms or photons inside the considered beam. The
average values of these operators are defined with the normalized
transverse wave-function $\psi_{\perp}$, following usual
conventions in quantum mechanics. For instance, the second-order
moment $\langle \hat{r}^2 \rangle$ reads
\begin{equation}
 \langle \hat{r}^2 \rangle (u)= \int {\rm d} x  {\rm d} y
 \psi_{\bot}^*(x,y,u)  (x^2+y^2)   \psi_{\bot}(x,y,u) \,.
\nonumber
\end{equation}
 Let us
note that, because of the correspondence $2k \rightarrow \hbar$,
the Hamiltonian~\eqref{eq:paraxial Hamiltonian} does not have the
dimension of an energy. Accordingly, we shall refer to the three
operators $\hat{K},\hat{V},\hat{U}$ as ``effective potentials''.
$\hat{K}$ is proportional to the transverse kinetic energy, while
$\hat{V}$ and $\hat{U}$ play respectively the roles of an
effective 2D-contact potential and of an external potential. The proposed
analysis of the propagation essentially relies on the commutation
relations between these three operators.

\section{Free nonlinear propagation of optical and atomic beams.}

We treat here the simpler case for which the paraxial
equation~\eqref{eq:paraxial equation} contains no effective
quadratic potential. This corresponds to the propagation of an
optical beam in a Kerr medium of uniform linear index, or to that
of an interacting atom laser in the absence of external potential.
We use the hidden symmetry of the paraxial equation to derive the
equations of motion for the moment $\langle \hat{r}^2 \rangle$
associated with such beams. We obtain a simple expansion law,
which we apply to discuss experimental results~\cite{David08a} on
guided atom lasers.

\subsection{Effective Hamiltonian scale invariance and propagation of the second-order moment.}

 The Hamiltonian~\eqref{eq:paraxial Hamiltonian} associated
with the free propagation equation reduces to
\begin{equation}
\hat{H}_0=\epsilon \hat{K}+\hat{V} \,.
\end{equation}
 The key point is to
investigate how this operator transforms in a space dilatation
$\mathbf{r} \rightarrow \lambda \mathbf{r}$~\cite{Pitaevskii97}
\begin{eqnarray}
\mathbf{r} \rightarrow \lambda \mathbf{r}, \: \: \psi(\mathbf{r})
\rightarrow \lambda^{d/2} \psi(\mathbf{r}/\lambda), \: \: \hat{H}_0
\rightarrow -\frac {\epsilon} {\lambda^2} \Delta_{T} + V \left(
\lambda \mathbf{r} \right) \nonumber \,.
\end{eqnarray}
To ensure the scale invariance of the Hamiltonian $\hat{H}_0$, the
interaction potential should transform as $V( \lambda \mathbf{r} )
= \lambda^{-2} V( \mathbf{r} )$. This is verified by the potential
$V(\mathbf{r})=1/r^2$, but also by the 2D-contact potential
$V(\mathbf{r})=\frac 1 2 \gamma
\delta^{(2)}(\mathbf{r})$~\cite{KaganPitaev}. The
 scale invariance of $\hat{H}_0$ can be expressed equivalently in
 terms of commutation with the generator $\hat{Q}$ of dilatations
 as
\begin{equation}
\label{eq:scale invariance commutator}
 [\hat{Q}, \hat{H}_0]= 2 i \hat{H}_0 \,.
\end{equation}
This generator reads $\hat{Q}= \frac {1} {2 \hbar} \left(
\hat{\mathbf{p}} \cdot \hat{\mathbf{r}} + \hat{\mathbf{r}} \cdot
\hat{\mathbf{p}} \right)$, it operates at each point~$\mathbf{r}$
a translation proportional to the vector~$\mathbf{r}$. The
evolution of the second-order moment (identified to the width
$w^2= \langle \hat{r}^2 \rangle$ for a light beam) is given by
Ehrenfest's theorem, applied with the
Hamiltonian~\eqref{eq:paraxial Hamiltonian}
\begin{equation}
\label{eq:optical parabolic moments} \frac {{\rm d} \langle
\hat{r}^2 \rangle} {{\rm d} u} = \frac {2 \epsilon} {k} \langle
\hat{Q} \rangle, \quad \frac {{\rm d}^2 \langle \hat{r}^2 \rangle}
{{\rm d} u^2} = \frac {\epsilon} {i k^2} \langle  [\hat{Q},
\hat{H}_0 ] \rangle \,.
\end{equation}
 The parabolic evolution of
$\langle \hat{r}^2 \rangle$ is indeed a direct consequence of the
Hamiltonian scale invariance~\eqref{eq:scale invariance
commutator}: the quantity ${\rm d}^2 \langle \hat{r}^2 \rangle /
{\rm d}u^2$ is proportional to the average effective energy
$\langle \hat{H}_0  \rangle$ of the transverse wave-function
\begin{equation} \frac {{\rm d}^2 \langle \hat{r}^2 \rangle} {{\rm d} u^2} =
\frac {2 \epsilon} { k^2} \langle   \hat{H}_0  \rangle, \nonumber
\end{equation}
which is a constant of motion. This result holds only for a 2D
transverse space, the only dimensionality giving the desired scale
transformation of the contact potential. In this case, using
Eq.\eqref{eq:scale invariance commutator} and Eq.\eqref{eq:optical
parabolic moments}, one obtains readily the free expansion law
valid for optical and atomic beams
\begin{equation}
\label{eq:parabolic law optical atomic beams} \langle \hat{r}^2
\rangle(u)= \frac {\epsilon} { k^2} \langle \hat{H}_0  \rangle \:
u^2 +  \frac {2 \epsilon} {k} \langle \hat{Q} \rangle_0 \: u +
\langle \hat{r}^2 \rangle_0 \,.
\end{equation}
 We noted $\langle \quad \rangle_0$ the average value of
the considered operator at the initial time of the propagation,
namely for the coordinate $u=0$.

\subsection{Application to the time-of-flight analysis of an atom laser beam.}

The previous result can be extended easily to account for the
presence of a linear gravitational potential in the propagation of
an atomic beam.  By considering the free-fall frame, or
alternatively by applying Ehrenfest's theorem as in
Eqs.~\eqref{eq:optical parabolic moments} with an additional
potential $V_g=2 k^2 g y$ (with $O_y$ the vertical axis), one
retrieves a parabolic law of Eq.\eqref{eq:parabolic law optical
atomic beams} for the transverse width $w^2(\tau)=\langle
\hat{x}^2 \rangle-\langle \hat{x} \rangle^2+\langle \hat{y}^2
\rangle-\langle \hat{y} \rangle^2$. We consider a wave-function
corresponding to a centered atomic beam, verifying initially
$\langle \hat{\mathbf{r}} \rangle_0=0$ and $\langle \hat{Q}
\rangle_0=0$. $\epsilon=1$ and $k= m / \hbar$ in the atomic
equation, so the
 free atomic Hamiltonian is $\hat{H}_0= \hat{K}+\hat{V}$. The width evolution reads
\begin{equation}
\label{eq:parabolic coefficient}  w^2(\tau)=  \frac {2} {k^2}
\left( \langle \hat{K} \rangle_0 +  \langle \hat{V} \rangle_0
\right) \tau^2+w^2(0) \,.
\end{equation}
The quantity $\langle \hat{K} \rangle_0= m^2 / \hbar^2 \langle
\Delta v^2 \rangle_{0}$ reflects the contribution of the
transverse kinetic energy to the beam divergence, while the
quantity $\langle \hat{V} \rangle_0$ reflects that of mean-field
interactions. This second contribution is usually discarded in the
time-of-flight analysis of atom laser beams.

The expansion law~\eqref{eq:parabolic coefficient} can be readily
applied to investigate interactions effects on the transverse
dynamics of a guided atom laser. Assuming that one knows
sufficiently the initial wave-function profile as to estimate
$\langle \hat{V} \rangle_0$, this equation gives the transverse
velocity dispersion $\langle \Delta v^2 \rangle_{0}$ as a function
of the width expansion $w(\tau)$ measured experimentally
\begin{equation}
\label{eq:definition delta vint} \langle \Delta v^2 \rangle_{0} =
\frac {1} {2 \tau^2} \left( w^2(\tau)- w^2(0) - \hbar^2  / m^2
\langle \hat{V} \rangle_0 \right) \nonumber \,.
\end{equation}
Interaction effects have been neglected in the analysis of the
expansion given in~\cite{Guerin06,David08a}, in which the
expansion is simply attributed to the initial transverse kinetic
energy. This leads to a different estimate of the transverse
velocity dispersion $\langle \Delta v^2 \rangle_{free}$, expressed
as a function of the beam width
\begin{equation}
\label{eq:definition delta vfree} \langle \Delta v^2
\rangle_{free} = \frac {1} {2  \tau^2} \left( w^2(\tau)- w^2(0)
\right) \,. \nonumber
\end{equation}
Discarding interaction effects in the beam expansion thus yields
the following error on the transverse velocity
\begin{equation}
\label{eq:calcul erreur sur deltavint} \frac {\langle \Delta v^2
\rangle_{free}- \langle \Delta v^2 \rangle_{0}} {\langle \Delta
v^2 \rangle_{0}} = \frac {\langle \hat{V} \rangle_0}  {\langle
\hat{K} \rangle_0} \,.
\end{equation}
For the quasi-monomode beam reported in~\cite{David08a}, one can
estimate the ratio $\langle \hat{V} \rangle_0/\langle \hat{K}
\rangle_0$ by modelling the initial transverse wave-function with
a Gaussian profile verifying $\langle \mathbf{r} \rangle_0=\langle
\mathbf{p} \rangle_0=0$~\footnotemark[2]\footnotetext[2]{To
evaluate this error for a multiple-mode atomic
beam~\cite{Guerin06}, one needs to know the mode decomposition of
the transverse wave profile in order to estimate $\langle \hat{V}
\rangle_0$.}. Eq.\eqref{eq:calcul erreur sur deltavint} shows then
that the value $\langle \Delta v^2 \rangle_{free}$ overestimates
the correct square velocity dispersion $\langle \Delta v^2
\rangle_{0}$ as $\langle \Delta v^2 \rangle_{free}=(1+2 n_{1D}
a_s)\langle \Delta v^2 \rangle_{0}$, thus of roughly $18\%$
in~\cite{David08a} with the reported linear density. In fact, interaction effects may
indeed affect significantly the flight of the atomic cloud, and
can only be safely neglected in the limit $n_{1D} a_s \ll
1$~\cite{Menotti02}.

\section{$ABCD$ matrix analysis of the nonlinear propagation of optical and atomic beams in presence of a quadratic potential.}

We now address the nonlinear propagation described by the
 full Eq.~\eqref{eq:paraxial equation} and
obtain an exact $ABCD$ matrix analysis of the beam evolution. As
seen previously, the additional effective potential contained in
this equation reflects either a position-dependent linear
permittivity of the form $\epsilon_r(x,y,z)=\epsilon^0_{r}(1-
\beta(z)^2 r^2)$ for an optical beam, either a time-independent
quadratic potential $U_{\perp}(r,z)= m/2 \times
\omega_{\perp}^2(z) \: r^2$ for an atomic beam. We remind that,
although they are assumed time-independent, the coefficient
$\beta$ and the trapping frequency $\omega_{\perp}$ may depend on
the longitudinal coordinate~$z$ or parameter~$\tau$. The presented treatment
can thus address atom laser experiments focusing the beam thanks
to a time-independent quadratic transverse potential which
sharpens along the propagation axis.

We proceed as follows. Using commutation relations between the
three operators $\hat{K},\hat{V},\hat{U}$ defined
in~Eqs.\eqref{eq:paraxial Hamiltonian terms}, we derive the
equations of motion for the second-order moment. This approach is
inspired from the treatment of Ghosh~\cite{Ghosh01}, but indeed
these equations had been derived several times
before~\cite{Vlasov71,Porras93,Pare92,Perez99}. We then introduce
a nonlinear quality factor for matter-wave beams, the invariance
of which is related to the hidden symmetry of the wave equation.
This parameter enables the definition of a nonlinear complex
radius of curvature $q$ ``renormalized'' by the mean-field
interactions. We show that this radius of curvature can be
propagated through $2 \times 2$ $ABCD$ matrices identical to those
used for linear optical and atomic beams. This conclusion agrees
with the results obtained by Porras \textit{et
al.}~\cite{Porras93} in nonlinear optics. To ease the comparison
between the linear and nonlinear regimes, an introduction to the
linear $ABCD$ matrix formalism for optical and atomic beams is
given in Appendix.

The presented treatment avoids the limitations of previous nonlinear $ABCD$ methods developed for matter waves~\cite{Chen08,Impens08b}.
In the non-paraxial $ABCD$ approach~\cite{Impens08b}, the
interaction term is considered as a perturbation, assumption which
becomes invalid for dense clouds or for long propagation times.
The paraxial $ABCD$ method used so far for interacting atomic
beams~\cite{Chen08,Impens08b}, inspired from
Ref.~\cite{Belanger83}, requires the absence of external potential
and, most critically, assumes a Gaussian-shape approximation of
the beam likely to break down after a finite propagation time.
 The advantage of the proposed treatment is that it addresses interacting atomic
beams in a fully non-perturbative manner. As in previous nonlinear
paraxial $ABCD$ approaches~\cite{Chen08,Impens08b}, this
method applies only to guided atomic beams experiencing no
significant longitudinal potential and propagating at a constant
longitudinal velocity. This hypothesis is realistic for the guided
atom lasers reported in the Refs.\cite{Guerin06,David08a}.

\subsection{Solutions for the propagation in a transverse quadratic potential.}

In the presence of a quadratic potential $\hat{U}$, the
Hamiltonian~\eqref{eq:paraxial Hamiltonian} $\hat{H}=\epsilon
\hat{K}+\hat{V}+\epsilon \hat{U}$ is no longer scale-invariant,
but the operators $\hat{H}_0=\epsilon \hat{K}+
\hat{V},\hat{U},\hat{Q}$ verify nonetheless the special set of
commutation relations~\cite{Pitaevskii97}
\begin{equation}
\label{eq:commutations relations} [\hat{Q},\hat{H}_0]  =  2 i
\hat{H}_0 \quad [\hat{Q},\epsilon \hat{U}]= -2 i \epsilon \hat{U}
\quad [ \epsilon \hat{U}, \hat{H}_0 ] =  4 i \alpha^2 k^2 \hat{Q}
\,.
\end{equation}
Thanks to this set of equations, signature of a hidden symmetry,
the Hamiltonian can be embedded into an $SO(2,1)$
algebra~\cite{Pitaevskii97}. These relations yield, through
Ehrenfests's theorem applied with $\hat{H}=\hat{H}_0+\epsilon
\hat{U}$, a closed set of coupled equations for the derivatives of
the second-order moment
\begin{equation}
\label{eq:moments GRIN}
 \left\{ \begin{aligned}
 \frac {{\rm d} \langle \hat{r}^2 \rangle} {{\rm d} u} & =  \frac {2
\epsilon} {k} \langle \hat{Q} \rangle \\
 \frac {{\rm d}^2 \langle
\hat{r}^2 \rangle} {{\rm d} u^2} & =  \frac {2 \epsilon} {k^2}
\left(  \langle \hat{H}_0 \rangle- \epsilon \langle \hat{U}
\rangle \right) \\
\frac {{\rm d}^3 \langle \hat{r}^2 \rangle} {{\rm d} u^3}  & = - 4
\alpha^2
 \frac {{\rm d} \langle \hat{r}^2 \rangle} {{\rm d} u} - 2
 \frac {{\rm d} \alpha^2} {{\rm d} u} \langle \hat{r}^2 \rangle \,.
\end{aligned} \right.
\end{equation}
Eqs.\eqref{eq:moments GRIN} show that, for a constant coefficient
$\alpha(u)=\alpha_0$ (or uniform trap frequency
$\omega_{\bot}(z)=\omega_0$), the second-order moment oscillates
with a frequency independent from the strength of the
nonlinearity. This property of nonlinear optics, established
in~\cite{Pare92}, is formally identical to the universal
low-energy mode found for 2D condensates~\cite{Pitaevskii97}.
Exact solutions may be found with a periodic quadratic coefficient
$\alpha(u)$ (or periodically modulated frequency
$\omega_{\bot}(\tau)$ for atomic beams) by noticing that the width
obeys a Hill's equation~\cite{Perez99,Ghosh01}.

\subsection{Nonlinear Quality Factor and Radius of curvature.}

The proposed $ABCD$ law applies to the propagation a parameter $q$
which can be interpreted as a complex radius of curvature
generalized for the nonlinear propagation. This parameter has been
introduced in previous developments of non-linear
optics~\cite{Pare92,Porras93}, and here we extend its definition
to atom optics. This definition involves an invariant of propagation, obtained from the set of
commutation relations~\eqref{eq:commutations relations} behind the
hidden symmetry, which reads
\begin{equation}
\label{eq:quality factor} M_{I}^4 = \epsilon \langle \hat{r}^2
\rangle \langle \hat{H}_0 \rangle - \langle \hat{Q} \rangle^2 \,.
\end{equation}
 For the
optical propagation in which $\epsilon=-1$, one retrieves the
nonlinear-optics quality factor~\cite{Siegman91,Porras93}. The
connection of this optics invariant with the hidden symmetry of
the paraxial equation has never been pointed out to my knowledge.
For the atomic propagation in which $\epsilon=1$, the parameter
$M_I$ constitutes the generalization to paraxial interacting beams
of the quality factor introduced by Riou~\textit{et
al.}~\cite{Riou06} for dilute atomic beams. The term $\langle
\hat{H}_0 \rangle$ includes in $M_I$ the additional divergence
resulting from interaction effects, which were not accounted for
in the quality factor defined in Ref.~\cite{Riou06}. The interacting
quality factor of a cylindrical and centered ($\langle \mathbf{r}
\rangle=0$) atomic beam in the fundamental Gaussian mode reads
$M_{I}^4=1+2 n_{1D} a_s$. In the limit of dilute atomic waves, one
retrieves $M_I=1$, so the interacting quality factor coincides
with the parameter defined in~\cite{Riou06}. It is also useful to
introduce the radius of curvature $R$~\cite{Pare92,Porras93}
\begin{equation}
\label{eq:radius curvature} \frac {1} {R} = \frac {1} {2 w^2}
\frac {{\rm d} w^2} {{\rm d} u} \,.
\end{equation}
With these parameters at hand, the generalized complex radius of
curvature reads
\begin{equation}
\label{eq:new complex radius curvature} \frac {1} {q} = \frac {1}
{R} + \frac {i M_I^2} {k w^2} \,.
\end{equation}

\subsection{Demonstration of the $ABCD$ law for the nonlinear propagation.}

The previous assumptions regarding the uniform linear atomic
density and cylindrical symmetry of the effective transverse
potential still apply. Unless specified otherwise, we also assume
that the beam is centered ($\langle \hat{\mathbf{r}} \rangle=0$)
and that the potential contains no linear term~\footnotemark[3]
\footnotetext[3]{In presence of a quadratic potential, a linear
gravitational potential can nonetheless be easily accounted for
with a change of coordinates $x'=x, \: y'=y+g / \omega_{\perp}^2$
(sag), and by considering the evolution of the moment $\langle
\hat{r}'^2 \rangle$ instead of $\langle \hat{r}^2 \rangle$.}.

Considering the radii of curvature $q_1=q(0)$ and $q_2=q(u)$ at
different stages of the propagation, we seek to establish an $ABCD$
law of the form
\begin{equation}
\label{eq:ABCDlaw}
 q_2= \frac {A q_1+B} {C q_1+D}  \,.
\end{equation}
It is sufficient that the two following relations be
satisfied~\cite{Porras93}
\begin{eqnarray}
\label{eq:input output w2} w_2^2 & = & w_1^2 \left[ A+ \frac {B}
{R_1} \right]^2+ \frac
{  M_I^4 B^2} {k^2 w_1^2}  \\
 \frac {w_2^2} {R_2} & = &  w_1^2 \left[ A+ \frac {B} {R_1} \right] \left[ C+ \frac {D} {R_1} \right]+ \frac
{  M_I^4 BD} {k^2 w_1^2} \label{eq:input output w2 sur R} \,.
\end{eqnarray}
The two functions $F_1(u)$ and $F_2(u)$, defined respectively by
the RHS of Eqs.~(\ref{eq:input output w2}) and (\ref{eq:input
output w2 sur R}), obey the same first-order differential system
as the functions $w^2$ and $w^2/R$ -obtained by considering the
set of Eqs.~\eqref{eq:moments GRIN} and Eqs.~(\ref{eq:quality
factor},\ref{eq:radius curvature})-
\begin{equation}
\frac {{\rm d} F_1} {{\rm d} u}  =  2 F_2, \quad \quad  \frac
{{\rm d} F_2} {{\rm d} u} F_1 = \frac {M^4_I} {k^2} + F_2^2 -
\alpha^2(u) F_1^2 \,.
\end{equation}
if and only if the $ABCD$ matrix satisfies the differential
equation
\begin{equation}
\label{eq:differential ABCD}
\frac {{\rm d}} {{\rm d} u} \left( \begin{array} {cc} A & B \\
                          C & D  \end{array} \right) = \left( \begin{array} {cc} 0 & 1 \\
                           - \alpha^2(u) & 0  \end{array}
                          \right) \left( \begin{array} {cc} A & B \\
                          C & D  \end{array} \right) \,,
\end{equation}
as well as the condition $(AD-BC)(u)=1$ for any parameter $u$.
Setting the initial condition $[A,B,C,D](0)=[1,0,0,1]$ guarantees
then that $F_1(u)$ and $F_2(u)$ coincide with $w^2$ and $w^2/R$ at
all times, thereby ensuring the $ABCD$ law~\eqref{eq:ABCDlaw}. The
nonlinear $ABCD$ matrix obeys Eq.\eqref{eq:differential ABCD}, which
is also the differential equation satisfied by the $ABCD$ matrix
describing the linear propagation of dilute atomic beams[See
Appendix Eq.~\eqref{eq:equation ABCD matrix linear atomic}]. This
shows that in spite of the mean-field interactions, the usual
interaction-free paraxial $ABCD$ matrices~\cite{Riou08} apply
rigorously to propagate the parameter $q$. In particular, one
retrieves the matrix
\begin{equation}
\left( \begin{array} {cc} A & B \\
                          C & D  \end{array} \right) = \left( \begin{array} {cc} 1 & u \\
                          0 & 1  \end{array} \right) \nonumber
\end{equation}
 for the free propagation~\footnotemark[4]\footnotetext[4]{For the free propagation, this result can be
obtained directly by noting that Eqs.\eqref{eq:optical parabolic
moments} imply $(1/q)'=-1/q^2$, and thus simply
$q(\tau)=q(0)+\tau$. By replacing $\langle \hat{r}^2 \rangle$ with
$w^2$ in the definition~\eqref{eq:quality factor} of the quality
factor, this relation can be extended to the propagation of a
non-centered beam ($\langle \hat{\mathbf{r}} \rangle \neq 0$) in a
linear potential.} and the matrix
\begin{equation}
\left( \begin{array} {cc} A & B \\
                          C & D  \end{array} \right) = \left( \begin{array} {cc}
                          \cos(\alpha u)  & \alpha^{-1} \sin(\alpha u) \\
                          - \alpha \sin(\alpha u)  & \cos(\alpha u)  \end{array} \right) \nonumber
\end{equation}
for the propagation in a quadratic index or cylindrical potential
 described by Eq.\eqref{eq:paraxial equation}. These matrices are identical to those
 given in the
Eqs.(\ref{eq:linear optics ABCD free},\ref{eq:linear optics ABCD
quadratic}) for the linear optical propagation, or in the
Eqs.(\ref{eq:linear atom optics ABCD free},\ref{eq:linear atom
optics ABCD quadratic}) for the linear atom-optical propagation. It
is a rather remarkable fact that the linear $ABCD$ laws persists for
an interacting beam: the inclusion of the invariant $M_I$ in the
definition of the radius of curvature $q$ captures the nonlinear
effects on the propagation. This is indeed a consequence of the
special set of commutation relations resulting from the hidden
symmetry.

Setting $u=\tau$ and $\alpha=\omega_{\perp}$ in these matrices,
one obtains the $ABCD$ law for the propagation of an interacting
atomic beam. An analogous property for the propagation of the
parameter $q$ had been established in a more restrictive
framework: it applied only to free-propagating paraxial Gaussian
atomic beams treated with a Gaussian-shape
approximation~\cite{Chen08,Impens08b}, which implies a
perturbative treatment of the interaction-term. The result
obtained here is much more general: it is non-perturbative, it
applies to atomic beams propagating in transverse potentials, and
it does not require any approximation once the assumptions of
paraxiality, uniform linear density of the atomic beam (and thus
absence of a longitudinal potential), and cylindrical symmetry of
the transverse potential are satisfied.

 In a similar manner as in~\cite{Pare92,Porras93}, this approach could also be
used to discuss the self-trapping in future experimental beams
involving atoms with attractive interactions, such as fermions
with Feshbach-tuned interactions. Self-trapping is expected at a
critical density yielding $M_I=0$, i.e. $n_{1D}= 1/(2 |a_s|)$ for
Gaussian beams, beyond which the beam collapses.

\section{Conclusion.}

To summarize, I have exploited the symmetries of the Hamiltonian
associated with the nonlinear paraxial wave equation - scale
invariance in the free propagation, specific set of commutation
relations in presence of a quadratic external potential - to extract
simple propagation laws for the width of a paraxial atomic beam of
uniform density. These results can be applied to analyse the
divergence of atom laser beams in recent
experiments~\cite{Guerin06,David08a} and evaluate their number of
transverse modes. A quality factor [Eq.~\eqref{eq:quality factor}]
has been defined, valid for paraxial interacting atomic beams
propagating in cylindrical potentials, thereby generalizing the
parameter introduced in~\cite{Riou06} for dilute paraxial atom
lasers. This parameter, together with a generalized radius of
curvature, allows one to describe with linear ABCD matrices the
propagation of a paraxial interacting atomic beam in constant,
cylindrical and quadratic potentials. An interesting question is the
robustness of the presented approach to a relaxation of the hypothesis of
paraxiality and uniform density of the atomic beam. Recent
theoretical work~\cite{Impens08b} on a matter-wave
resonator~\cite{Impens06b}, involving a free-falling atomic cloud
bouncing on curved atomic mirrors, shows that the width oscillations
of universal frequency persist with a good approximation in spite of
a time-dependent linear density. This suggests that this treatment
could be fruitfully applied to guided atom lasers presenting a slow
variation of this parameter. Last, let us point out a strong
connection between the physics of guided atom lasers and of 2D
condensates, involving a similar equation. We have seen that
theoretical tools for the latter can also be relevant for nonlinear
atom optics. This also suggests that several effects specific of 2D
condensates, such as the breathing modes, might be reproduced within
the transverse profile of guided atom laser beams.

The author thanks David Gu\'ery-Odelin and Yann Le Coq for fruitful discussions,
William Gu\'erin for comments, and Steve Walborn
for manuscript reading. He acknowledges a very rich collaboration on atom optics
with Christian Bord\'e, and thanks Luiz Davidovich and Nicim Zagury for hospitality. This work was supported
by DGA(Contract No 0860003), by the Ecole Polytechnique and by the French Ministry of Foreign Affairs (Lavoisier-Brésil Grant).

\appendix

\section{The $ABCD$ matrix formalism for linear paraxial optics and atom optics.}
\label{app:ABCD formalism for linear optics atom optics}

This Appendix gives an introduction to the $ABCD$ matrix formalism
for optical and atomic beams experiencing a linear propagation.

\subsection{The $ABCD$ formalism in linear paraxial optics.}

The $ABCD$ matrix formalism, discussed in various
textbooks~\cite{Siegman}, was originally introduced to analyse the
light wave propagation in linear optical systems involving a
constant or piece-wise quadratic index of
refraction~\cite{Kogelnik65}. We present briefly this approach for
ray and Gaussian optics.

One associates to a light ray a two-component vector map involving
the ray distance to the axis $r(z)$ and the slope $r'(z)= {\rm
d}r(z)/ {\rm d}z$. In a linear optical system, the propagation of
this vector is given by a linear input-output relation of the form
\begin{equation}
\label{eq:general ABCD law} \left( \begin{array} {c} r(z_2) \\
r'(z_2)
\end{array} \right) = \left( \begin{array} {cc} A & B \\ C & D
\end{array} \right) \left(
\begin{array} {c} r(z_1) \\ r'(z_1) \end{array} \right) \,,
\end{equation}
called $ABCD$ law, in which the matrix coefficients depend on the
propagation distance $\Delta z=z_2-z_1$, and satisfy the relation
$AD-BC=1$. In a medium of uniform index, the $ABCD$ matrix is
simply
\begin{equation}
\label{eq:linear optics ABCD free}
 \left( \begin{array} {cc} A & B \\ C & D
\end{array} \right)=
\left( \begin{array} {cc} 1 & \Delta z \\ 0 & 1 \end{array}
\right) \,.
\end{equation}
In a graded index medium with $n(r,z)=n_0(1-1/2\beta^2(z) r^2)$,
the ray propagation equation yields the matrix
\begin{equation}
\label{eq:linear optics ABCD quadratic}
\left( \begin{array} {cc} A &   B   \\
C & D
\end{array} \right) =\left( \begin{array} {cc} \mbox{cos}(\beta \Delta z) &   \beta^{-1}\mbox{sin}(\beta \Delta z)   \\
-\beta \mbox{sin}(\beta \Delta z) & \mbox{cos}(\beta \Delta z)
\end{array} \right) \,.
\end{equation}

The $ABCD$ formalism is especially well-suited to propagate
Gaussian waves. Let us consider an input beam at the plane $z=z_1$
with a transverse profile
\begin{equation}
\label{eq:Gaussian optical mode} \psi_{\bot}(x,y,z_1)= \mbox{exp}
\left( - i \frac {k x^2} {2 q_1} \right)  \quad \mbox{with} \quad
\frac {1} {q_1} = \frac {1} {R_1} - \frac {2 i} { k w_1^2} \, .
\end{equation}
 $R_1$ is the radius of curvature and $w_1$ the beam width in the direction
 $O_x$, $q_1$ is called the complex radius of curvature. The quantity $k=n_0 \omega / c $ is the longitudinal
wave-vector of the optical wave. The beam is assumed to propagate
between the planes $z=z_1$ and $z=z_2=z_1+\Delta z$ in a linear
optical system characterized by an $ABCD$ matrix. The beam
propagator in this system is directly related to the $ABCD$ matrix
elements~\cite{Siegman}
\begin{equation}
\label{eq:paraxial optics propagator} K(x,x',\Delta z)=\sqrt{\frac
{i k} {2 \pi B} } \exp \left( - \frac {i k} {2 B} ( A x'^2 -2x' x+
D x^2 ) \right) \,.
\end{equation}
 The output beam at $z=z_2$ is given by the
propagation integral
\begin{equation}
\label{eq:ABCDoptics propagation integral} \psi_{\bot}(x,y,z_2)=
\int_{-\infty}^{+\infty} {\rm d}x' K(x',x,z_2-z_1)
\psi_{\bot}(x',y,z_1) \,,
\end{equation}
and can be expressed as
\begin{equation}
\psi_{\bot}(x,y,z_2)= \sqrt{\frac {1} {A+ B/ q_1}} \mbox{exp}
\left( - i \frac {k x^2} {2 q_2} \right) \nonumber \,,
\end{equation}
with the complex radius of curvature $q_2$ given by the $ABCD$
law, rewritten here for convenience
\begin{equation}
\label{eq:ABCDlaw appendix} q_2 = \frac {A q_1 + B} {C q_1  + D}
\,.
\end{equation}
 This treatment can be extended to the full Hermite-Gauss basis~\cite{Siegman}.

\subsection{The $ABCD$ formalism in linear paraxial atom optics.}

Let us now apply the $ABCD$ matrix technique to propagate dilute
paraxial atomic waves~\cite{LeCoq01,Riou08}. The corresponding
propagation equation has the form of
Eq.\eqref{eq:atomlaserparaxial0} without the interaction term
($n_{1D}\simeq0$)
\begin{equation}
i \hbar \frac {\partial \psi_{\bot}} {\partial \tau}= \left[-\frac
{\hbar^2} {2m} \Delta_T + V_x(x,\tau) \right] \psi_{\bot} \,.
\end{equation}
The potential $V_x$ is assumed to be quadratic of the form $V_x=
m \omega_{\perp}^2 x^2/2$. The propagator $K$ is given by the Van
Vleck formula or by the $ABCD$ formalism~\cite{BordeHouches}
\begin{equation}
\label{eq:paraxial atom optics propagator} K(x,x',\tau)=
\sqrt{\frac {i m} { 2 \hbar \pi  B} } \exp \left(  \frac {i m} {2
\hbar B} ( A x'^2 -2 x' x+ D x^2 ) \right) \,.
\end{equation}
If one sets $k=m/\hbar$ as previously for atomic beams, this
expression becomes formally identical to Eq.\eqref{eq:paraxial
optics propagator} up to a sign change in the argument of the
exponential. The corresponding $ABCD$ matrix describes the
one-dimensional motion of a classical particle moving in the
potential $V_x$. Precisely, let us consider a classical particle
initially at the position $x(0)$ and with a speed $v_x(0)$, moving
in the potential $V_x$. At the instant $t=\tau$, this particle has
a final position $x(\tau)$ and a speed $v_x(\tau)$. Since the
equations of motion are linear, the final values are related to
the initial ones by a linear map analogous to Eq.\eqref{eq:ABCDlaw
appendix}
\begin{equation}
\label{eq:ABCD matrix Hamilton} \left( \begin{array} {c} x(\tau) \\
v_x(\tau)
\end{array} \right) = \left( \begin{array} {cc} A(\tau) & B(\tau)
\\ C(\tau) & D(\tau)
\end{array} \right) \left(
\begin{array} {c} x(0) \\ v_x(0) \end{array} \right) \,.
\end{equation}
The $ABCD$ matrix involved in this relation is precisely the
matrix used in the propagator of Eq.\eqref{eq:paraxial atom optics
propagator}. Let us verify this. One can check that the
distribution defined by Eq.\eqref{eq:paraxial atom optics
propagator} satisfies the Schr\"odinger equation if and only if
the coefficients $A,B,C,D$ satisfy the first-order differential
equation
\begin{equation}
\label{eq:equation ABCD matrix linear atomic} \frac {\rm d} {\rm d
\tau} \left(
\begin{array} {cc} A & B
\\ C & D
\end{array} \right) = \left( \begin{array} {cc} 0 & 1 \\ -\omega^2_{\perp} & 0 \end{array} \right)
\left( \begin{array} {cc} A & B \\ C & D
\end{array} \right) \,.
\end{equation}
It is easy to show that the $ABCD$ matrix defined in
Eq.\eqref{eq:ABCD matrix Hamilton} obeys the same equation.
Besides, when the time $\tau$ tends towards $0$, the $ABCD$ matrix
defined by Eq.\eqref{eq:ABCD matrix Hamilton} tends towards
identity. In this limit, the distribution defined with this matrix
in Eq.\eqref{eq:paraxial atom optics propagator} yields a Dirac
distribution, as expected for a propagator between equal instants.
Eq.~\eqref{eq:paraxial atom optics propagator} and the $ABCD$
matrix of Eq.\eqref{eq:ABCD matrix Hamilton} thus define a
distribution $K$ which satisfies the Schr\"odinger equation as
well as the correct initial condition. This shows that $K$ is the
propagator of the Schr\"odinger equation.

 Using Eq.\eqref{eq:equation ABCD matrix linear atomic}, it is straightforward to establish the expression of the $ABCD$ matrix
for the free motion
\begin{equation}
\label{eq:linear atom optics ABCD free}
\left( \begin{array} {cc} A & B \\
C & D
\end{array}
 \right)= \left( \begin{array} {cc} 1 & \tau \\
0 & 1
\end{array}
 \right) \,,
\end{equation}
and for the motion in a quadratic potential
\begin{equation}
\label{eq:linear atom optics ABCD quadratic}
\left( \begin{array} {cc} A & B \\
C & D
\end{array}
 \right)=\left( \begin{array} {cc} \mbox{cos}(\omega_{\perp} \tau) &  \omega_{\perp}^{-1} \mbox{sin}(\omega_{\perp} \tau) \\
- \omega_{\perp} \mbox{sin}(\omega_{\perp} \tau) &
\mbox{cos}(\omega_{\perp} \tau)
\end{array}
 \right) \,.
\end{equation}

The expression of the propagator can be readily applied to obtain,
by analogy with Gaussian optics, an $ABCD$ law for the propagation
of Gaussian wave-functions. Let us consider the complex radius of
curvature, decomposed as previously along its imaginary and
complex part as
\begin{equation}
\frac {1} {q_{1}} = \frac {1} {R_{1}} + \frac {2 i } {k w_1^2} \,.
\nonumber
\end{equation}
We consider the initial transverse wave-function $\psi_{\perp}$,
initially of the form
\begin{equation} \label{eq:spherical wave
function} \psi_{\perp} ( x, y, \tau_1 ) = \left( \frac {2} {\pi}
\right)^{1/4} \frac {1} {\sqrt{w_1}} \exp \left[ i \frac {k} {2
q_{1}} x^2 \right] \,.
\end{equation}
Note that this wave-function is properly normalized to unity. Its
evolution is given by the propagation integral
\begin{equation}
\label{eq:propagation integral} \psi_{\bot}(x,\tau_2)= \int {\rm
d} x' K(x,x',\tau) \psi_{\bot}(x',y',\tau_1) \,, \nonumber
\end{equation}
with $\tau=\tau_2-\tau_1$. Using the optical
integral~\eqref{eq:ABCDoptics propagation integral}, one obtains
the wave-function at the instant $\tau_2$
\begin{equation}
 \psi_{\perp} ( x, y, \tau_2 ) = \left( \frac {2} {\pi}
\right)^{1/4} \frac {1} {\sqrt{w_2}} \exp \left[ i \frac {k} {2
q_{2}} x^2 \right] \,, \nonumber
\end{equation}
where $q_{1}$ and $q_{2}$ satisfy the $ABCD$ law of
Eq.\eqref{eq:ABCDlaw appendix}.

This formalism can be readily extended to the propagation of
wave-functions in a separable quadratic transverse potential
$V_{\perp}=V_x+V_y$. The corresponding propagator is simply the
product of the one-dimensional propagators
$K(x,x',y,y',\tau)=K_x(x,x',\tau)K_y(y,y',\tau)$ defined by
Eq.\eqref{eq:paraxial atom optics propagator}. A $2 \times 2$
$ABCD$ matrix is associated with each direction. The propagation
of a Gaussian wave-function
\begin{equation} \psi_{\perp} ( x, y, \tau_1 ) = \sqrt{\frac {2} {\pi w_{1x} w_{1y}}} \exp \left[ i \frac {k} {2 q_{1x}} x^2 \right] \exp \left[ i
\frac {k} {2 q_{1y}} y^2 \right]  \nonumber
\end{equation}
is formally identical: the parameters $q_x$ and $q_y$ satisfy a
relation of the form of Eq.\eqref{eq:ABCDlaw appendix} with the
$ABCD$ matrices associated with their respective directions.


\begin{thebibliography}{45}
\expandafter\ifx\csname
natexlab\endcsname\relax\def\natexlab#1{#1}\fi
\expandafter\ifx\csname bibnamefont\endcsname\relax
  \def\bibnamefont#1{#1}\fi
\expandafter\ifx\csname bibfnamefont\endcsname\relax
  \def\bibfnamefont#1{#1}\fi
\expandafter\ifx\csname citenamefont\endcsname\relax
  \def\citenamefont#1{#1}\fi
\expandafter\ifx\csname url\endcsname\relax
  \def\url#1{\texttt{#1}}\fi
\expandafter\ifx\csname
urlprefix\endcsname\relax\def\urlprefix{URL }\fi
\providecommand{\bibinfo}[2]{#2}
\providecommand{\eprint}[2][]{\url{#2}}


\bibitem[{\citenamefont{Impens}(2008)}]{AtomLaserTheory}
\bibinfo{author}{\bibfnamefont{C.~J.} \bibnamefont{Bord\'{e}}},
  \bibinfo{journal}{Phys. Lett. A} \textbf{\bibinfo{volume}{204}},
  \bibinfo{pages}{217} (\bibinfo{year}{1995});  \bibinfo{author}{\bibfnamefont{M.}~\bibnamefont{Holland}},
 \bibinfo{author}{\bibfnamefont{K.}~\bibnamefont{Burnett}},
  \bibinfo{author}{\bibfnamefont{C.}~\bibnamefont{Gardiner}},
   \bibinfo{author}{\bibfnamefont{J.~I.}~\bibnamefont{Cirac}},
   and  \bibinfo{author}{\bibfnamefont{P.}~\bibnamefont{Zoller}},
  \bibinfo{journal}{Phys. Rev. A} \textbf{\bibinfo{volume}{54}},
  \bibinfo{pages}{R1757} (\bibinfo{year}{1996});
  \bibinfo{author}{\bibfnamefont{A.~M.} \bibnamefont{Guzman}},
\bibinfo{author}{\bibfnamefont{M.} \bibnamefont{Moore}},
\bibinfo{author}{\bibfnamefont{P.} \bibnamefont{Meystre}}
  \bibinfo{journal}{Phys. Rev. A} \textbf{\bibinfo{volume}{53}},
  \bibinfo{pages}{977} (\bibinfo{year}{1996}); \bibinfo{author}{\bibfnamefont{H.~M.} \bibnamefont{Wiseman}},
\bibinfo{author}{\bibfnamefont{A.-M.} \bibnamefont{Martins}},
and \bibinfo{author}{\bibfnamefont{D.~F.} \bibnamefont{Walls}},
  \bibinfo{journal}{Quantum Semiclassical Opt.} \textbf{\bibinfo{volume}{8}},  \bibinfo{pages}{737}
  (\bibinfo{year}{1996}); \bibinfo{author}{\bibfnamefont{H.~M.} \bibnamefont{Wiseman}},
  \bibinfo{journal}{Phys. Rev. A} \textbf{\bibinfo{volume}{56}},
  \bibinfo{pages}{2068} (\bibinfo{year}{1997}).

\bibitem[{\citenamefont{Impens}(2008)}]{AtomLaserExp}
\bibinfo{author}{\bibfnamefont{M.-O.} \bibnamefont{Mewes\textit{ et~al.}}},  \bibinfo{journal}{Phys. Rev. Lett.}
\textbf{\bibinfo{volume}{78}},
  \bibinfo{pages}{582} (\bibinfo{year}{1997}); \bibinfo{author}{\bibfnamefont{B.~P.}
  \bibnamefont{Anderson}} and \bibinfo{author}{\bibfnamefont{M.~A.}
  \bibnamefont{Kasevich}},  \bibinfo{journal}{Science}
\textbf{\bibinfo{volume}{282}},
  \bibinfo{pages}{1686} (\bibinfo{year}{1998}); \bibinfo{author}{\bibfnamefont{E.~W.}
  \bibnamefont{Hagley\textit{ et~al.}}} \bibinfo{journal}{Science}
  \textbf{\bibinfo{volume}{283}}, \bibinfo{pages}{1706 }
  (\bibinfo{year}{1999});
  \bibinfo{author}{\bibfnamefont{I.}~\bibnamefont{Bloch}},
  \bibinfo{author}{\bibfnamefont{T.~W.}~\bibnamefont{H\"{a}nsch}},
   and \bibinfo{author}{\bibfnamefont{T.}~\bibnamefont{Esslinger}},
  \bibinfo{journal}{Phys. Rev. Lett.} \textbf{\bibinfo{volume}{82}},
  \bibinfo{pages}{3008} (\bibinfo{year}{1999}); \bibinfo{author}{\bibfnamefont{N. P.}~\bibnamefont{Robins\textit{ et~al.}}},
  \bibinfo{journal}{Phys. Rev. Lett.} \textbf{\bibinfo{volume}{96}},
  \bibinfo{pages}{140403} (\bibinfo{year}{2006})

\bibitem[{\citenamefont{LeCoq\textit{ et~al.}}(2001)}]{LeCoq01}
\bibinfo{author}{\bibfnamefont{Y.}~\bibnamefont{LeCoq\textit{ et~al.}}},
  \bibinfo{journal}{Phys. Rev. Lett.} \textbf{\bibinfo{volume}{87}},
  \bibinfo{pages}{170403} (\bibinfo{year}{2001}).

\bibitem[{\citenamefont{Guerin\textit{ et~al.}}(2006)}]{Guerin06}
\bibinfo{author}{\bibfnamefont{W.}~\bibnamefont{Guerin\textit{ et~al.}}},
  \bibinfo{journal}{Phys. Rev. Lett.} \textbf{\bibinfo{volume}{97}},
  \bibinfo{pages}{200402} (\bibinfo{year}{2006}).

\bibitem[{\citenamefont{Couvert\textit{ et~al.}}(2008)}]{David08a}
\bibinfo{author}{\bibfnamefont{A.}~\bibnamefont{Couvert\textit{ et~al.}}},
  \bibinfo{journal}{Eur. Phys. Lett. \textbf{\bibinfo{volume}{83}}, 50001 }  (\bibinfo{year}{2008}).






\bibitem{Interfero} \bibfnamefont{C.~J.} \bibnamefont{Bord\'{e}}
\bibinfo{journal}{Phys. Lett. A} \textbf{\bibinfo{volume}{140}}
\bibinfo{pages}{10} (\bibinfo{year}{1989}); \bibinfo{author}{\bibfnamefont{P.}~\bibnamefont{Bouyer}} \bibnamefont{and}
  \bibinfo{author}{\bibfnamefont{M.A.}~\bibnamefont{Kasevich}},
  \bibinfo{journal}{Phys. Rev. A} \textbf{\bibinfo{volume}{56}},
  \bibinfo{pages}{R1083} (\bibinfo{year}{1997});
  \bibinfo{author}{\bibfnamefont{S.}~\bibnamefont{Gupta}},
 \bibinfo{author}{\bibfnamefont{K.}~\bibnamefont{Dieckmann}},
  \bibinfo{author}{\bibfnamefont{Z.}~\bibnamefont{Hadzibabic}},
   \bibinfo{author}{\bibfnamefont{D.~E.}~\bibnamefont{Pritchard}},
  \bibinfo{journal}{Phys. Rev. Lett.} \textbf{\bibinfo{volume}{89}},
  \bibinfo{pages}{140401} (\bibinfo{year}{2002}); \bibinfo{author}{\bibfnamefont{Y.-J.} \bibnamefont{Wang\textit{ et~al.}}},
  \bibinfo{journal}{Phys. Rev. Lett.} \textbf{\bibinfo{volume}{94}},
  \bibinfo{pages}{090405} (\bibinfo{year}{2005});
  \bibinfo{author}{\bibfnamefont{Y.}~\bibnamefont{LeCoq}},
\bibinfo{author}{\bibfnamefont{J.~A.}~\bibnamefont{Retter}},
\bibinfo{author}{\bibfnamefont{S.}~\bibnamefont{Richard}},
\bibinfo{author}{\bibfnamefont{A.}~\bibnamefont{Aspect}},
and \bibinfo{author}{\bibfnamefont{P.}~\bibnamefont{Bouyer}},
  \bibinfo{journal}{Appl. Phys. B: Lasers Opt.}
  \textbf{\bibinfo{volume}{84}}, \bibinfo{pages}{627}
  (\bibinfo{year}{2006}).

  \bibitem{Tobias}
  \bibinfo{author}{\bibfnamefont{T.}~\bibnamefont{Paul}},
\bibinfo{author}{\bibfnamefont{K.}~\bibnamefont{Richter}},
and \bibinfo{author}{\bibfnamefont{P.}~\bibnamefont{Schlagheck}},
  \bibinfo{journal}{Phys. Rev. Lett.} \textbf{\bibinfo{volume}{94}},
  \bibinfo{pages}{020404} (\bibinfo{year}{2005}{\natexlab{a}});
   \bibinfo{author}{\bibfnamefont{T.}~\bibnamefont{Paul}},
\bibinfo{author}{\bibfnamefont{P.}~\bibnamefont{Leboeuf}},
\bibinfo{author}{\bibfnamefont{N.}~\bibnamefont{Pavloff}},
\bibinfo{author}{\bibfnamefont{K.}~\bibnamefont{Richter}},
and \bibinfo{author}{\bibfnamefont{P.}~\bibnamefont{Schlagheck}},
  \bibinfo{journal}{Phys. Rev. A} \textbf{\bibinfo{volume}{72}},
  \bibinfo{pages}{063621} (\bibinfo{year}{2005}{\natexlab{b}}). %T. Ernst, T. Paul and P. Schlagheck, e-print arXiv:0905.4750 (2009).


\bibitem{LeboeufPavloff} \bibinfo{author}{\bibfnamefont{P.}~\bibnamefont{Leboeuf}} \bibnamefont{and}
  \bibinfo{author}{\bibfnamefont{N.}~\bibnamefont{Pavloff}},
  \bibinfo{journal}{Phys. Rev. A} \textbf{\bibinfo{volume}{64}},
  \bibinfo{pages}{033602} (\bibinfo{year}{2001}); \bibinfo{author}{\bibfnamefont{N.}~\bibnamefont{Pavloff}},
  \bibinfo{journal}{Phys. Rev. A} \textbf{\bibinfo{volume}{66}},
  \bibinfo{pages}{013610} (\bibinfo{year}{2002}).


\bibitem{DavidMoments} D. Gu\'ery-Odelin, F. Zambelli, J. Dalibard, and S. Stringari,
Phys. Rev. A \textbf{60}, 4851 (1999);
\bibinfo{author}{\bibfnamefont{D.}
\bibnamefont{Gu\'ery-Odelin}}, in \textit{Peyresq Lectures on
Nonlinear Phenomena}, edited by J-A Sepulchre (World
Scientific, Singapore, 2003),  Vol. II.

\bibitem{KaganPitaev} \bibinfo{author}{\bibfnamefont{L.}~\bibnamefont{Pitaevskii}},
  \bibinfo{journal}{Phys. Lett. A} \textbf{\bibinfo{volume}{221}},
  \bibinfo{pages}{14} (\bibinfo{year}{1996}); Yu. Kagan, E. L. Surkov, and G. V. Shlyapnikov
  \bibinfo{journal}{Phys. Rev. A} \textbf{\bibinfo{volume}{54}},
  \bibinfo{pages}{R1753} (\bibinfo{year}{1996}).

\bibitem[{\citenamefont{Pitaevskii and Rosch}(1997)}]{Pitaevskii97}
\bibinfo{author}{\bibfnamefont{L.~P.} \bibnamefont{Pitaevskii}}
  \bibnamefont{and} \bibinfo{author}{\bibfnamefont{A.}~\bibnamefont{Rosch}},
  \bibinfo{journal}{Phys. Rev. A} \textbf{\bibinfo{volume}{55}},
  \bibinfo{pages}{R853} (\bibinfo{year}{1997}).

\bibitem[{\citenamefont{Vlasov\textit{ et~al.}}(1971)}]{Vlasov71}
 S.~N.Vlasov, V.A.Petrishchev, and V.I.Talanov
  \bibinfo{journal}{Radiophys. Quant. Electr.} \textbf{\bibinfo{volume}{14}},
  \bibinfo{pages}{1062} (\bibinfo{year}{1971}).


\bibitem[{\citenamefont{Pare and Belanger}(1992)}]{Pare92}
\bibinfo{author}{\bibfnamefont{C.}~\bibnamefont{Par\'e}} \bibnamefont{and}
  \bibinfo{author}{\bibfnamefont{P.}~\bibnamefont{B\'elanger}},
  \bibinfo{journal}{Opt. and Quant. Electr.}
  \textbf{\bibinfo{volume}{24}}, \bibinfo{pages}{S1051} (\bibinfo{year}{1992}).



\bibitem[{\citenamefont{Porras\textit{ et~al.}}(1993)}]{Porras93}
M. Porras, J. Alda, and E. Bernabeu,
  \bibinfo{journal}{Appl. Opt.} \textbf{\bibinfo{volume}{32}},
  \bibinfo{pages}{5885} (\bibinfo{year}{1993}).




\bibitem[{\citenamefont{Bord\'{e}}(1991)}]{BordeHouches}
C. J. Bord\'e, in \textit{Fundamental Systems in Quantum Optics}, Les
Houches Lectures LIII (Elsevier, New York, 1991).


\bibitem[{\citenamefont{Lenz\textit{ et~al.}}(1993)}]{Lenz93}
G. Lenz, P. Meystre, and E. M. Wright,
  \bibinfo{journal}{Phys. Rev. Lett.} \textbf{\bibinfo{volume}{71}},
  \bibinfo{pages}{3271} (\bibinfo{year}{1993}).

\bibitem{NonlinearAtomexp} \bibinfo{author}{\bibfnamefont{L.}~\bibnamefont{Deng\textit{ et~al.}}},
  \bibinfo{journal}{Nature} \textbf{\bibinfo{volume}{398}},
  \bibinfo{pages}{218} (\bibinfo{year}{1999})
; \bibinfo{author}{\bibfnamefont{S.}~\bibnamefont{Burger\textit{ et al.}}},
  \bibinfo{journal}{Phys. Rev. Lett.} \textbf{\bibinfo{volume}{83}},
  \bibinfo{pages}{5198} (\bibinfo{year}{1999}); \bibinfo{author}{\bibfnamefont{S.}~\bibnamefont{Inouye\textit{ et~al.}}},
  \bibinfo{journal}{Nature} \textbf{\bibinfo{volume}{402}},
  \bibinfo{pages}{641} (\bibinfo{year}{1999}{\natexlab{a}}); \bibinfo{author}{\bibfnamefont{S.}~\bibnamefont{Inouye\textit{ et~al.}}},
  \bibinfo{journal}{Science} \textbf{\bibinfo{volume}{285}},
  \bibinfo{pages}{571} (\bibinfo{year}{1999}{\natexlab{b}}); \bibinfo{author}{\bibfnamefont{J.}~\bibnamefont{Denschlag\textit{ et~al.}}},
  \bibinfo{journal}{Science} \textbf{\bibinfo{volume}{287}},
  \bibinfo{pages}{97} (\bibinfo{year}{2000}); \bibinfo{author}{\bibfnamefont{L.}~\bibnamefont{Khaykovich\textit{ et al.}}},
  \bibinfo{journal}{Science} \textbf{\bibinfo{volume}{296}},
  \bibinfo{pages}{1290} (\bibinfo{year}{2002}); \bibinfo{author}{\bibfnamefont{K.~E.} \bibnamefont{Strecker\textit{ et al.}}},
  \bibinfo{journal}{Nature} \textbf{\bibinfo{volume}{417}}, 150
  (\bibinfo{year}{2002}).




\bibitem[{\citenamefont{Meystre}(2001)}]{MeystreBook}
\bibinfo{author}{\bibfnamefont{P.}~\bibnamefont{Meystre}},
  \emph{\bibinfo{title}{Atom Optics}} (\bibinfo{publisher}{Springer Verlag},
  \bibinfo{year}{2001}).




\bibitem[{\citenamefont{Kogelnik}(1965)}]{Kogelnik65}
\bibinfo{author}{\bibfnamefont{H.}~\bibnamefont{Kogelnik}},
  \bibinfo{journal}{Bell Sys. Tech. Journ.} \textbf{\bibinfo{volume}{44}},
  \bibinfo{pages}{455} (\bibinfo{year}{1965}).

\bibitem{BordeABCD} \bibinfo{author}{\bibfnamefont{C.~J.} \bibnamefont{Bord\'{e}}},
  \bibinfo{journal}{C. R. Acad. Sci. Paris} \textbf{\bibinfo{volume}{4}},
  \bibinfo{pages}{509} (\bibinfo{year}{2001}{\natexlab{a}}); \bibinfo{author}{\bibfnamefont{C.~J.} \bibnamefont{Bord\'{e}}},
  \bibinfo{journal}{Metrologia} \textbf{\bibinfo{volume}{39}},
  \bibinfo{pages}{435} (\bibinfo{year}{2002}{\natexlab{b}}).

\bibitem[{\citenamefont{Riou\textit{ et~al.}}(2008)}]{Riou08}
\bibinfo{author}{\bibfnamefont{J.-F.} \bibnamefont{Riou\textit{ et~al.}}},
  \bibinfo{journal}{Phys. Rev. A} \textbf{\bibinfo{volume}{77}},
  \bibinfo{pages}{033630} (\bibinfo{year}{2008}).


\bibitem[{\citenamefont{Riou\textit{ et~al.}}(2006)}]{Riou06}
\bibinfo{author}{\bibfnamefont{J.-F.} \bibnamefont{Riou\textit{ et~al.}}},
  \bibinfo{journal}{Phys. Rev. Lett.} \textbf{\bibinfo{volume}{96}},
  \bibinfo{pages}{070404} (\bibinfo{year}{2006}).




\bibitem[{\citenamefont{Impens}(2008)}]{Impens08a}
\bibinfo{author}{\bibfnamefont{F.}~\bibnamefont{Impens}},
  \bibinfo{journal}{Phys. Rev. A} \textbf{\bibinfo{volume}{77}},
  \bibinfo{pages}{013619} (\bibinfo{year}{2008}).

\bibitem[{\citenamefont{Busch\textit{ et~al.}}(2002)}]{Busch02}
T. Busch, M. K\"{o}hl, T. Esslinger, and K. M\o lmer,
  \bibinfo{journal}{Phys. Rev. A} \textbf{\bibinfo{volume}{65}},
  \bibinfo{pages}{043615} (\bibinfo{year}{2002}).



\bibitem{DavidGuidedPropagation}
\bibinfo{author}{\bibfnamefont{T.} \bibnamefont{Lahaye\textit{ et~al.}}},
  \bibinfo{journal}{Comm. Nonlinear Sci. Numer. Simul.} \textbf{\bibinfo{volume}{8}},
  \bibinfo{pages}{315} (\bibinfo{year}{2003}).


\bibitem[{\citenamefont{Ghosh}(2001)}]{Ghosh01}
\bibinfo{author}{\bibfnamefont{P.~K.} \bibnamefont{Ghosh}},
  \bibinfo{journal}{Phys. Rev. A} \textbf{\bibinfo{volume}{65}},
  \bibinfo{pages}{012103} (\bibinfo{year}{2001}).


\bibitem{AppliNonlinOptics} \bibinfo{author}{\bibfnamefont{C.}~\bibnamefont{Par\'e}} \bibnamefont{and}
  \bibinfo{author}{\bibfnamefont{P.}~\bibnamefont{B\'elanger}},
  \bibinfo{journal}{IEEE Journ. Quant. Elec.} \textbf{\bibinfo{volume}{30}},
  \bibinfo{pages}{1141} (\bibinfo{year}{1994}); \bibinfo{author}{\bibfnamefont{P.}~\bibnamefont{B\'elanger}} \bibnamefont{and}
  \bibinfo{author}{\bibfnamefont{N.}~\bibnamefont{B\'elanger}},
  \bibinfo{journal}{Opt. Commun.} \textbf{\bibinfo{volume}{117}},
  \bibinfo{pages}{56} (\bibinfo{year}{1995});  \bibinfo{author}{\bibfnamefont{S.}~\bibnamefont{Nemoto}},
  \bibinfo{journal}{Appl. Opt.} \textbf{\bibinfo{volume}{34}},
  \bibinfo{pages}{6123} (\bibinfo{year}{1995}); V. M. P\'erez-Garc\'{\i}a, P. Torres, J. J. Garc\'{i}a-Ripoll and
H. Michinel, J. Opt. B: Quantum Semiclassical Opt. \textbf{2}, 353
(2000).


\bibitem{BookNonLinOptics} Guang S. He, Song H. Liu, \textit{Physics of Non Linear Optics}, (World Scientific, Singapore
1999).


\bibitem{Jackson03}
A. D. Jackson,  G. M. Kavoulakis, and C. J. Pethick,
  \bibinfo{journal}{Phys. Rev. A} \textbf{\bibinfo{volume}{58}},
  \bibinfo{pages}{2417} (\bibinfo{year}{1998}).



\bibitem{Menotti02}
\bibinfo{author}{\bibfnamefont{C.}~\bibnamefont{Menotti}} \bibnamefont{and}
  \bibinfo{author}{\bibfnamefont{S.}~\bibnamefont{Stringari}},
  \bibinfo{journal}{Phys. Rev. A} \textbf{\bibinfo{volume}{66}},
  \bibinfo{pages}{043610} (\bibinfo{year}{2002}).


\bibitem[{\citenamefont{Impens and Borde}(2007)}]{Impens08b}
\bibinfo{author}{\bibfnamefont{F.}~\bibnamefont{Impens}} \bibnamefont{and}
  \bibinfo{author}{\bibfnamefont{C.~J.} \bibnamefont{Bord\'e}},
  Phys. Rev. A \textbf{\bibinfo{volume}{79}},
  \bibinfo{pages}{043613} (\bibinfo{year}{2009}).


\bibitem{Chen08}
J. Chen, Z. Zhang, Y. Liu, and Q. Lin,  \bibinfo{journal}{Opt.
Expr.} \textbf{\bibinfo{volume}{16}},
  \bibinfo{pages}{10918} (\bibinfo{year}{2008}).


\bibitem[{\citenamefont{B\'{e}langer and \'{e}}(1983)}]{Belanger83}
\bibinfo{author}{\bibfnamefont{P.-A.} \bibnamefont{B\'{e}langer}}
  \bibnamefont{and} \bibinfo{author}{\bibfnamefont{C.} \bibnamefont{Par\'{e}}},
  \bibinfo{journal}{Applied Optics} \textbf{\bibinfo{volume}{22}},
  \bibinfo{pages}{1293} (\bibinfo{year}{1983}).









\bibitem{Perez99} J. J. Garc\'{i}a-Ripoll, V. M. P\'erez-Garcia and P.
Torres, Phys. Rev. Lett. \textbf{\bibinfo{volume}{83}},
  \bibinfo{pages}{1715} (\bibinfo{year}{1999}).



\bibitem{Siegman91} \bibinfo{author}{\bibfnamefont{A. E.}~\bibnamefont{Siegman}},
  \bibinfo{journal}{IEEE J. Quantum Electron.} \textbf{\bibinfo{volume}{27}},
  \bibinfo{pages}{1146} (\bibinfo{year}{1991}).




\bibitem{Impens06b}
F. Impens, P. Bouyer and C. J. Bord\'e, \bibinfo{journal}{Appl.
Phys. B: Lasers Opt.} \textbf{\bibinfo{volume}{84}},
  \bibinfo{pages}{603} (\bibinfo{year}{2006}); \bibfnamefont{K.~J.}~\bibnamefont{Hughes},
\bibinfo{author}{\bibfnamefont{J.~H.~T.}~\bibnamefont{Burke}}, and
\bibinfo{author}{\bibfnamefont{C.~A.}~\bibnamefont{Sackett}}, Phys.
Rev. Lett. \textbf{\bibinfo{volume}{102}},
  \bibinfo{pages}{150403} (2009); F. Impens and C.~J. Bord\'e, Phys. Rev. A \textbf{80}, 031602(R) (2009).



\bibitem{Siegman} \bibinfo{author}{\bibfnamefont{A. E.}~\bibnamefont{Siegman}},
  \bibinfo{booktitle}{Lasers}, (Ed. University Science Books, 1990).

\end{thebibliography}
\end{document}